# Lung Nodule Detection in Screening Computed Tomography

Ilaria Gori, Roberto Bellotti, Piergiorgio Cerello, Sorin Cristian Cheran, Giorgio De Nunzio, Maria Evelina Fantacci, Parnian Kasae, Giovanni Luca Masala, Alessandro Preite Martinez and Alessandra Retico

*Abstract*–A computer-aided detection (CAD) system for the identification of pulmonary nodules in low-dose multi-detector helical Computed Tomography (CT) images with 1.25 mm slice thickness is presented. The basic modules of our lung-CAD system, a dot-enhancement filter for nodule candidate selection and a neural classifier for false-positive finding reduction, are described. The results obtained on the collected database of lung CT scans are discussed.

## I. INTRODUCTION

LUNG cancer is one of the most relevant public health issues. Despite significant research efforts and advances in the understanding of tumour biology, there was no reduction of the mortality over the last decades.

Lung cancer most commonly manifests itself with the formation of non-calcified pulmonary nodules. Computed Tomography (CT) is the best imaging modality for the detection of small pulmonary nodules [1], particularly since the introduction of the helical technology. However, the amount of data that need to be interpreted in CT examinations can be very large, especially when multi-detector helical CT and thin collimation are used, thus generating up to about 300 two-dimensional images per scan, corresponding to about 150 MB. In order to support radiologists in the identification of early-stage pathological objects, researchers have recently begun to explore computer-aided detection (CAD) methods in this area.

Among the approaches that are being tried to reduce the mortality of lung cancer is the implementation of screening programs for the subsample of the population with higher risk of developing the disease. The First Italian Randomized Controlled Trial that aims to study the potential impact of screening on a high-risk population using low-dose helical CT was recently started [2]. A CAD system for small pulmonary nodule identification, based on the analysis of images acquired in this trial, was developed in the framework of the MAGIC-5 collaboration funded by Istituto Nazionale di Fisica Nucleare (INFN) and Ministero dell'Università e della Ricerca (MIUR). The system is based on a dot-enhancement filter for the identification of nodule candidates and a neural network based classification module for the reduction of the number of false-positive (FP) findings per scan.

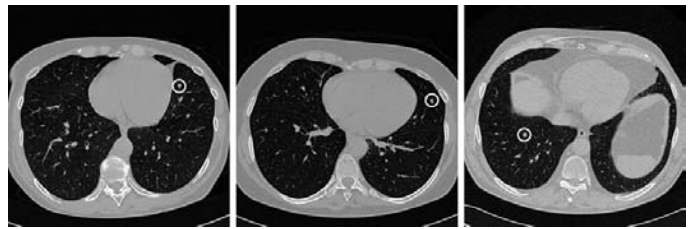

Fig. 1. Some examples of pulmonary nodules.

## II. THE CAD SYSTEM

Pulmonary nodules may be characterized by very low CT values and/or low contrast, may have CT values similar to those of blood vessels and airway walls or may be strongly connected to them (see Fig. 1).

The strategy we adopted focuses first on the detection of nodule candidates by means of a 3D enhancing filter emphasizing spherically-shaped objects. As a second step, the reduction of false-positive findings by means of a voxel-based neural approach is implemented. The two steps of the analysis are applied only on the lung volume identified by means of a segmentation algorithm that defines the internal region, according to the procedure proposed in [3].

The automated nodule candidate detection should be characterized by a sensitivity value close to 100%, in order to avoid setting an *a priori* upper bound to the CAD system performances. To this aim, lung nodules are modelled as spherical objects and a dot-enhancement filter is applied to the

Manuscript received November 17, 2006. This work was supported in part by Centro Studi e Ricerche Enrico Fermi and Bracco Imaging S.p.A..

I. Gori is with Bracco Imaging S.p.A., Milano, Italy and Istituto Nazionale di Fisica Nucleare, Sezione di Pisa, Pisa, Italy (e-mail: ilariagori@gmail.com).
R. Bellotti is with Dipartimento Interateneo di Fisica, Università di Bari, Bari, Italy, Istituto Nazionale di Fisica Nucleare, Sezione di Bari, Bari, Italy and TIRES - Center of Innovative Technologies for Signal Detection and Processing, Università di Bari, Bari, Italy (e-mail: roberto.bellotti@ba.infn.it).
P. Cerello is with Istituto Nazionale di Fisica Nucleare, Sezione di Torino, Torino, Italy (e-mail: cerello@to.infn.it).
S. C. Cheran is with Dipartimento di Fisica, Università di Genova, Genova, Italy and Istituto Nazionale di Fisica Nucleare, Sezione di Torino, Torino, Italy (e-mail: cheran@to.infn.it).
G. De Nunzio is with Dipartimento di Fisica, Università di Lecce, Lecce, Italy and Istituto Nazionale di Fisica Nucleare, Sezione di Lecce, Lecce, Italy (e-mail: giorgio.denunzio@unile.it).
M. E. Fantacci is with Dipartimento di Fisica, Università di Pisa, Pisa, Italy and Istituto Nazionale di Fisica Nucleare, Sezione di Pisa, Pisa, Italy (e-mail: fantacci@df.unipi.it).
P. Kasae is with Istituto Nazionale di Fisica Nucleare, Sezione di Cagliari, Cagliari, Italy (e-mail: parnian.kasae@ca.infn.it).
G. L. Masala is with Struttura Dipartimentale di Matematica e Fisica, Università di Sassari, Sassari, Italy and Istituto Nazionale di Fisica Nucleare, Sezione di Cagliari, Cagliari, Italy (e-mail: giovanni.masala@ca.infn.it).
A. Preite Martinez is with Centro Studi e Ricerche Enrico Fermi, Roma, Italy (e-mail: ale@incal.net).
A. Retico is with Istituto Nazionale di Fisica Nucleare, Sezione di Pisa, Pisa, Italy (e-mail: alessandra.retico@pi.infn.it).

3D matrix of voxel data. The filter attempts to determine the local geometrical characteristics of each voxel, by computing the eigenvalues of the Hessian matrix and evaluating a *magnitude* and a *likelihood* functions specifically configured to discriminate between the local morphology of linear, planar and spherical objects, the latest modelled as having 3D Gaussian sections [4], [5]. To enhance the sensitivity of this filter to nodules of different sizes, a multi-scale approach was followed. According to the indications given in [4],[6],[7], a Gaussian smoothing at several scales was implemented. In particular, the Gaussian smoothing and the computation of the Hessian matrix were combined in a convolution between the original data and the second derivatives of a Gaussian smoothing function. This procedure is based on an *a priori* knowledge of the size range of the nodules to be enhanced. The range and the number of intermediate smoothing scales must be determined empirically on the basis of the dataset of available CT scans. Once a set of N filtered images is computed, each voxel of the 3D space matrix is assigned the maximum *magnitude* x *likelihood* value obtained from the different scales, multiplied by the relative scale factor, according to [4], [7]. A peak-detection algorithm is then applied to the filter output to detect the local maxima in the 3D space matrix. The final filter output is a list of nodule candidates sorted by the value the filter function assigned. The list is empirically truncated so to keep the sensitivity close to 100%, therefore accepting a large number of false positives.

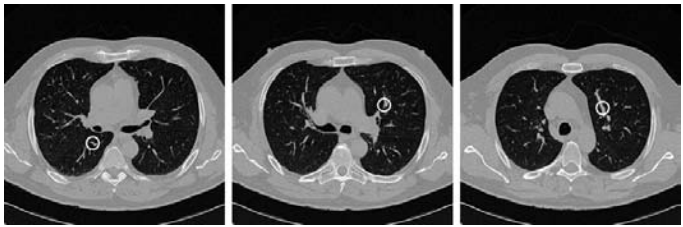

Fig. 2. Some examples of false positive findings generated by the dot-enhancement filter.

Most FP findings are crossings between blood vessels (see Fig. 2). To reduce the amount of FP/scan, we have developed a procedure we called voxel-based neural approach (VBNA). Each voxel of a region of interest (ROI) is characterized by a feature vector constituted by the grey level intensity values of its 3D neighbours (see Fig. 3) and the eigenvalues of the gradient and the Hessian matrices [8]. A feed-forward neural network is trained and tested at this stage assigning each voxel either to the nodule or normal tissue target class. A candidate nodule is then characterized as "CAD nodule" if the number of pixels within its ROI tagged as "nodule" by the neural classifier is above some relative threshold. A free response receiver operating characteristic (FROC) curve for our CAD system can therefore be evaluated at different threshold levels.

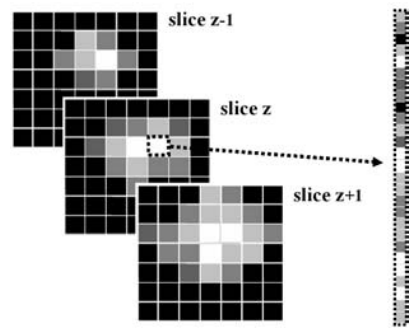

Fig. 3. Voxel-based neural approach to false-positive reduction.

## III. DATA ANALYSIS AND RESULTS

The CAD system was tested on a dataset of low-dose (screening setting: 140 kV, 70÷80 mA) CT scans with reconstructed slice thickness of 1.25 mm. The scans were collected and annotated by experienced radiologists in the framework of the screening trial being conducted in Tuscany (Italy) [2]. The database used for this study consists of 33 CT scans, containing 74 internal nodules with diameter between 5 and 12 mm. Each scan is a sequence of about 300 slices stored in DICOM (Digital Imaging and COmmunications in Medicine) format.

Once the internal region of the lung has been identified and the 3D dot-enhancement filter has provided the lists of nodule candidates, the VBNA is applied to reduce the amount of FP findings. The available dataset of 74 internal nodules was partitioned into a teaching set of 30 nodules belonging to 15 scans and a validation set of 44 nodules belonging to the other 18 scans; the partition was defined so that the teaching set is representative of all the nodule dimensions. Five three-layer feed-forward neural networks were trained on five different random partitions of the teaching set into train and test sets. The performances achieved in each trial for the correct classification of individual pixels are reported in Table I, where the sensitivity and the specificity values obtained on the test sets and on the whole teaching set in the five trials are shown.

TABLE I
VBNA: PERFORMACES ACHIEVED BY FIVE NEURAL NETWORKS ON FIVE RANDOM PARTITIONS OF TEACHING SET INTO TRAIN AND TEST SETS

| Test set | | Teaching set | |
|---|---|---|---|
| sens % | spec % | sens % | spec % |
| 78.2 | 83.8 | 85.6 | 86.0 |
| 80.0 | 82.8 | 87.5 | 86.6 |
| 77.2 | 85.3 | 81.7 | 86.2 |
| 74.0 | 82.3 | 85.6 | 85.9 |
| 77.6 | 79.3 | 84.8 | 81.0 |

The first three networks in Table I provided the best performances on test sets. Among them, the second one was more balanced with respect to sensitivity and specificity on the test set and achieved the best performance on the teaching set. So the VBNA approach has been applied to each ROI by using this trained neural network. The FROC curve (see Fig. 4) was evaluated on the whole dataset, constituted by the teaching set of 30 nodules belonging to 15 scans and the validation set of 44 nodules belonging to 18 scans. As shown in figure, a sensitivity of 86.5% at 6 FP/scan is measured. If the sensitivity value is decreased to 75.7% a rate of 3.8 FP/scan is obtained.

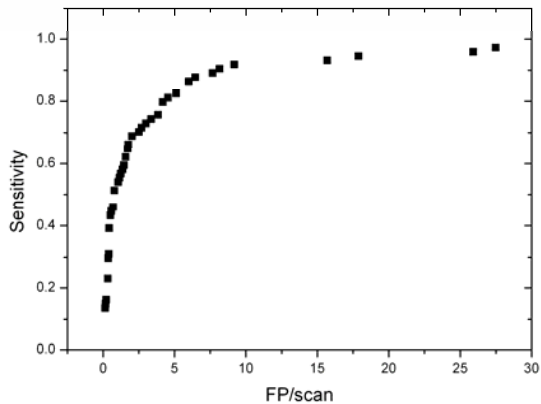

Fig. 4. FROC curve on the whole dataset of 74 internal nodules in 33 scans.

## IV. CONCLUSIONS

The dot-enhancement pre-processing is a suitable tool for the identification of nodule candidates of diameter above 5 mm and the VBNA is an effective approach to the problem of false positives reduction. In particular the VBNA approach provides a very good sensitivity (86.5%) at a low level of false positives (6); the sensitivity is still very high (75.7%) at less than 4 false positives per scan. The results obtained so far are promising, albeit a validation against a larger database is required.

ACKNOWLEDGMENT

We thank the researchers of the INFN- and MIUR-funded MAGIC-5 Collaboration for contributing to this work. We acknowledge Dr. L. Battolla, Dr. F. Falaschi and Dr. C. Spinelli of the U.O. Radiodiagnostica 2 dell'Azienda Ospedaliera Universitaria Pisana and Prof. D. Caramella and Dr. T. Tarantino of the Divisione di Radiologia Diagnostica e Interventistica del Dipartimento di Oncologia, Trapianti e Nuove Tecnologie in Medicina dell'Università di Pisa for providing the annotated database of CT scans. We are grateful to Dr M. Mattiuzzi from Bracco Imaging S.p.A. for useful discussions.